# Possible coexistence of s-wave and unconventional pairing in $Na_xCoO_2 \cdot yH_2O$: a new insight from impurity effects

Y.-J. Chen,[1] C.-J. Liu,[2] J.-S. Wang,[2] J.-Y. Lin,[1] C. P. Sun,[3] S. W. Huang,[4]
J. M. Lee,[5] J. M. Chen,[5] J. F. Lee,[5] D. G. Liu,[5] and H. D. Yang[3]

[1]*Institute of Physics, National Chiao Tung University, Hsinchu 300, Taiwan*
[2]*Department of Physics, National Changhua University of Education, Changhua 50007, Taiwan*
[3]*Department of Physics, National Sun Yat Sen University, Kaohsiung 804, Taiwan*
[4]*Department of Electrophysics, National Chiao Tung University, Hsinchu 300, Taiwan*
[5]*National Synchrotron Radiation Research Center, Hsinchu 300, Taiwan*

To shed light on the symmetry of the superconducting order parameter in $Na_xCoO_2 \cdot yH_2O$, the Mn doping effects are studied. X-ray absorption spectroscopy verifies that the doped Mn impurities occupy the Co sites and are with a valance close to +4. Impurity scattering by Mn is in the unitary limit that, however, does not lead to strong $T_c$ suppression. This absence of the strong impurity effects on $T_c$ is not consistent with the simple picture of a sign-changing order parameter. Coexistence of the *s*-wave and unconventional order parameters is proposed to reconcile all existing experiments and has been directly observed by the specific heat experiments.

$Na_xCoO_2 \cdot yH_2O$ has stimulated researchers due to its exotic superconductivity, the electronic frustration on the triangular Co lattice, and the characters of the strongly correlated electron systems. All of these subjects are of current interest in condensed matter physics [1,2,3]. There is no doubt that $Na_xCoO_2 \cdot yH_2O$ has been recognized as one of the most interesting superconductors since the high-$T_c$ superconductor era. Compared with $MgB_2$ in which the two-gap feature was concluded within one year of its discovery [4], $Na_xCoO_2 \cdot yH_2O$ appears to be more exotic judged by, for example, the proposed models of its superconductivity ([5-8] and references therein). The superconducting order parameter of $Na_xCoO_2 \cdot yH_2O$ has remained elusive. On the one hand, almost all the relevant experiments agree on the existence of the nodal lines in the order parameter [9-13] favoring *p*- or *f*-wave pairing, if the crystal and time reversal symmetries are further considered. On the other hand, the decrease in the Knight shift below $T_c$ [12,13] is a strong advocate of *s*- or *d*-wave pairing. To gain more insights into the nature of the superconductivity in $Na_xCoO_2 \cdot yH_2O$, the impurity scattering effects could provide one useful venue. Impurity effects on $T_c$ have revealed indispensable information about superconductivity in many novel superconductors such as cuprates [14-21] and $Sr_2RuO_4$ [22]. The general idea is that impurity potential scattering breaks Cooper pairs in the superconductors with a sign-changing order parameter, while it rarely affects $T_c$ of those with a fully gapped *s*-wave order parameter if the impurity doping does not change the carrier density significantly. To shed light on the crucial issue of the order parameter symmetry, $Na_{0.7}Co_{1-z}Mn_zO_2 \cdot yH_2O$ samples were prepared. For the first time, x-ray absorption spectroscopy (XAS) shows that the doped Mn ions indeed occupy the Co sites. However, the unitary impurity scattering by Mn does not lead to strong $T_c$ suppression. To reconcile the weak $T_c$ suppression by impurities with other experiments, coexistence of *s*-wave and the unconventional pairings is proposed. This model is further supported by the direct observation from the specific heat $C(T)$ experiments.

Polycrystalline parent compounds of sodium cobalt oxides $\gamma$-$Na_{0.7}Co_{1-z}Mn_zO_2$ ($z$ = 0 to 0.03) were prepared using a rapid heat-up procedure. High purity powders of $Na_2CO_3$, $Mn_2O_3$, and CoO were *thoroughly* mixed, ground, and calcined. The resulting powders were immersed in the 3M $Br_2/CH_3CN$ solution for 5 days, followed by filtering and thorough washing with $CH_3CN$ and DI water. X-ray diffraction (XRD) patterns indicated that all parent and hydrated samples were of single phase. To further characterize the samples, x-ray absorption spectroscopy (XAS) (including XANES and EXAFS) was carried out for Mn *K*- and *L*-edges as well as Co *K*-edge. Details of XAS experiments can be found in Refs. [23,24]. To ensure the accurate results, $T_c$ measurements were carried out, respectively, in two systems (MPMS and PPMS of Quantum Design). Results from two measurement systems and two different batches of samples were all consistent. Resistivity $\rho$ was measured by the standard four probe method. $C(T)$ was further measured to explore the phase transition at very low temperatures. A detailed description of the $C(T)$ measurements can be found in [25].

Indeed, there was at least one report on the Ir and Ga impurity effects in $Na_xCoO_2 \cdot yH_2O$ [26]. However, whether Ir or Ga impurities occupied the Co sites was not carefully examined. This uncertainty has undermined the persuasion of the previous work. To explore the impurity effects in cobaltates, $Na_xCo_{1-z}Mn_zO_2 \cdot yH_2O$ seems a natural choice. Layered $Alk_xMnO_2 \cdot yH_2O$ ($Alk$=Li, Na, K) with the triangular $MnO_2$ planes was successfully synthesized more than a decade ago [27,28]. This indicates that the doped Mn ions would favor the Co sites in $Na_xCoO_2 \cdot yH_2O$ and do not alter the crystal symmetry, as suggested by XRD. For further characterizations of the doped Mn ions, Fig. 1(a) shows Mn $L$-edge XANES spectra of $Na_xCo_{1-z}Mn_zO_2 \cdot yH_2O$ together with those of $MnO_2$, $Mn_2O_3$ and $MnO$. It is clear that all the Mn-doped samples have similar spectra distinct from those of the starting material $Mn_2O_3$ and two other common manganese oxides. Moreover, judged by the edge energy around 6550 eV in Fig. 1(b), Mn ions in all the doped samples, hydrated or not, have a valance close to +4. The $Mn^{4+}$ state is consistent with the previous suggestion that the valance of Mn in nominal $Alk_xMnO_2 \cdot yH_2O$ is close to +4 by the extraction reactions [27]. The first spectroscopic evidence of $Mn^{4+}$ is provided here for Mn ions on the triangular lattice. Fig. 1(c) shows the Fourier transform (FT) amplitudes and profiles of EXAFS data at Co and Mn $K$-edges for $Na_xCo_{0.99}Mn_{0.01}O_2 \cdot yH_2O$. FT amplitudes and profiles of EXAFS for both Co and Mn show identical features, indicating that Mn ions are in the same environment as Co ions and no other local structure of any second phase is present. FT peak position for the first shell of Co is 0.187 nm (Fig. 1(c)), identical to the Co-O bond length of $Na_xCoO_2 \cdot yH_2O$ from the neutron diffraction [29,30]. Also shown in Fig. 2(c), the Mn-O bond length is found to be 0.181 nm. This implies that the $Mn^{4+}$ ionic radius ($r_I$) of 0.052 nm might be smaller than that of the Co ions which valance is between +3.3 and +3.4 ($r_I$=0.065 nm for $Co^{3+}$; $r_I$ for $Co^{4+}$ not available) [31,32].

With the characterizations verifying that *the doped Mn impurities occupy the Co sites in $CoO_2$ planes*, one is ready to study the impurity effects on $T_c$ of $Na_xCo_{1-z}Mn_zO_2 \cdot yH_2O$. Fig. 2 shows $M$ vs. $T$ below 6 K for samples of the first batch. The inset demonstrates the Meissner effect of the undoped $Na_xCoO_2 \cdot yH_2O$, as pronounced as those of the best polycrystalline samples in the literature. $T_c$ is defined as the onset of the Meissner effect. The doping dependence of $T_c$ for two respective batches is depicted in Fig. 3. $T_c$ suppression rate $dT_c/dz$=0.64 K/1% is determined by fitting data of all samples in Fig. 2. For samples with $z$>0.02, albeit with a small drop in the $M$-$T$ curve, the signal is too small to well determine $T_c$ as discussed later.

The results in Fig. 3 are somewhat surprising. Previously, the nodal order parameter was detected in $Na_xCoO_2 \cdot yH_2O$. Therefore, $T_c$ should have been strongly suppressed by impurity scattering as in the cases of cuprates and $Sr_2RuO_4$. For example, $dT_c/dz$=12 K/1% in $YBa_2(Cu_{1-z}Zn_z)_3O_{6.93}$ [9-13]. To elucidate the impurity scattering due to Mn doping, $\rho(T)/\rho(T=245\ K)$ in Fig. 4(a) demonstrates the ratio of the carrier scattering rate at low $T$ (presumably dominated by impurity scattering) to that at 245 K, suggesting that Mn doping indeed leads to an increase in the impurity scattering rate for the parent compounds

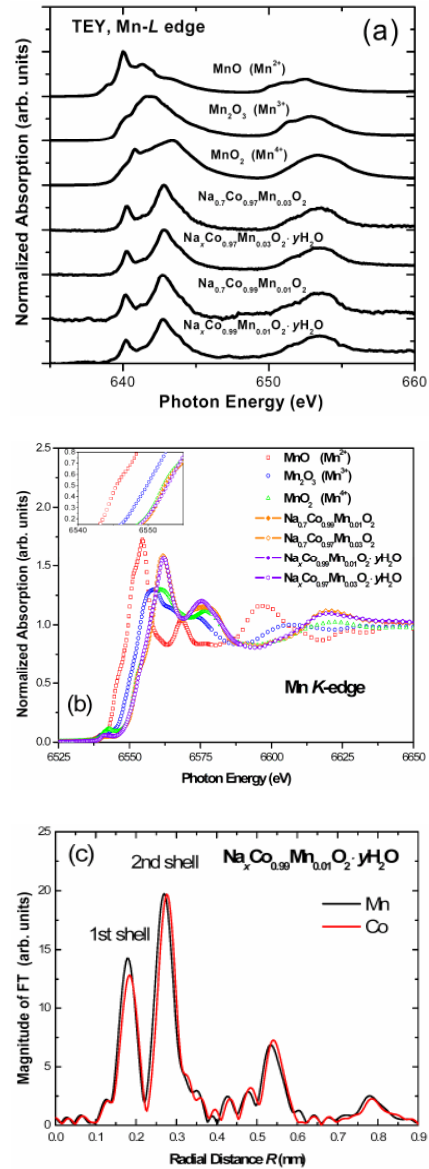

Fig. 1 Mn (a) $L$-edge and (b) $K$-edge XANES of $Na_xCo_{1-z}Mn_zO_2 \cdot yH_2O$ and the standard samples; (c) FT magnitudes with phase correction of the $k^3$-weighted EXAFS data at Mn $K$-edge. Inset of (b) shows the enlarged edge region.

$Na_{0.7}Co_{1-z}Mn_zO_2$. To discuss the impurity scattering in a more quantitative way, $\rho(T=10\ K)\equiv\rho_0$ vs. z can be seen in the inset of Fig. 4(b). Taking $d\rho_0/dz$=314 μΩ cm/1%, the increase in the sheet resistance $R$ of the $CoO_2$ planes due to impurities can be obtained by $R=\rho/(c/2)$ where $c$=0.55 nm is the $c$-axis of $Na_{0.7}CoO_2$. However, $\rho$ of polycrystalline samples does not necessarily represents the intrinsic value due to the complex current flow paths, grain boundaries, and other effects. Indeed, $\rho(T)$ of $Na_{0.7}CoO_2$ shown in Fig. 4(b) has a similar T dependence compared to that of the single crystal with the same composition, but with a five times larger magnitude [2]. Taking this factor of 5, $d\rho_0/dz$=63 μΩ cm/1% is the normalized value for $Na_{0.7}Co_{1-z}Mn_zO_2$, and consequently $dR_0/dz$=1100 Ω/1%. For two dimensional isotropic scattering, $R_0=4(\hbar/e^2)(z/n)\sin2\delta_0$, where $n$ is the carrier number and $\delta_0$ is the phase shift. In the unitary limit, $dR_0/dz$=545 $\Omega$/1% with a conservative value of $n$=0.3 per $CoO_2$ plane. Therefore, Mn impurities in $CoO_2$ planes indeed scatter carriers in the unitary limit. To compare to the Zn impurity scattering (usually considered as a case of the unitary limit) in $YBa_2(Cu_{1-z}Zn_z)_3O_{6.93}$, $d\rho_0/dz$=45 $\mu\Omega$ cm/1% From Ref. [15]. Assuming all Zn impurities go to the planes, $dR_0/dz$=528 $\Omega$/1%, and the theoretical unitary limit is $dR_0/dz$=1020 $\Omega$/1% with $n$=0.16 per $CuO_2$ plane. $T_c$ suppression rate for a two dimensional line nodal superconductor due to unitary scattering was calculated to be $dT_c/dz\approx$6 K/1% [14] and plotted in Fig. 3 as the dashed line.

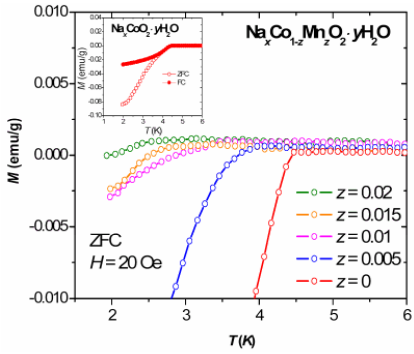

Fig. 2 $M$ vs. $T$ of $Na_xCo_{1-z}Mn_zO_2\cdot yH_2O$ samples. Inset shows ZFC and FC data of undoped $Na_xCoO_2\cdot yH_2O$.

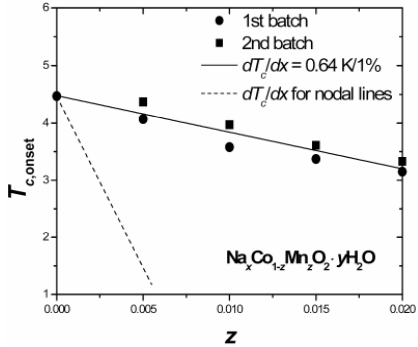

Fig. 3 $T_c$ suppression by Mn impurities. The solid line is the linear fit of all data. The observed $T_c$ suppression rate is much smaller than that (the dashed line) of a superconductor with line nodes (see text).

The slow $T_c$ suppression due to impurity scattering shown in Fig. 4 naively suggests a fully gapped $s$-wave superconductivity in $Na_xCoO_2\cdot yH_2O$. However, the existence of the nodal order parameter by other experiments hinders a simple $s$-wave scenario. In principle, the anisotropy of impurity scattering would lead to a weaker $T_c$ suppression for sign-changing order parameter. However, the observed $dT_c/dz$ in $Na_xCo_{1-z}Mn_zO_2\cdot yH_2O$ would require the ratio of anisotropic to isotropic scattering to be close to one, which is certainly unusual [18,21]. There are other proposals such as the triplet $s$-wave [6] and the extended $s$-wave [8] scenarios. With further consideration of the possible existence of the singlet state from NMR results, we propose that *two superconducting order parameters coexist in $Na_xCoO_2\cdot yH_2O$, which happen to have about the same $T_c$*. It is likely that the $s$-, $d$-, $f$-waves all have similar pairing strength, as in the calculations for a certain range of parameters [8]. In this coexistence model, the Mn impurities strongly suppress superconductivity with the nodal order parameter, while the $s$-wave component remains relatively robust. This model, though seemingly exotic, is one of the few that reconcile all three major facts: the existence of the nodal lines, that of the singlet state, and the weak $T_c$ suppression due to impurity scattering. Actually, specific heat data in Ref. [11], with further analysis, could accommodate the coexistence of a fully gapped (presumably $s$-wave) and a nodal order parameter with up to 60% of the Fermi surface gapped [33]. The minor $T_c$ suppression of the $s$-wave component by Mn impurities could be due to the change of the Co valance, which is known to have influence on $T_c$ and likely to be affected by doped $Mn^{4+}$ ions. Another possibility is the magnetic pair breaking by the spin-flip scattering rate $\tau_s^{-1}$ from the local moment of $Mn^{4+}$, which usually is one order of magnitude smaller than $\tau_{imp}^{-1}$ [34]. The signal of the Meissner effect becomes substantially weak for $z \geq 0.01$. This could be partly due to the suppression of superconductivity in the volume of the nodal order parameter. Moreover, impurities may perturb and remove the energy degeneracy of the two order parameters and the samples with increasing $z$ become favoring the nodal order parameter.

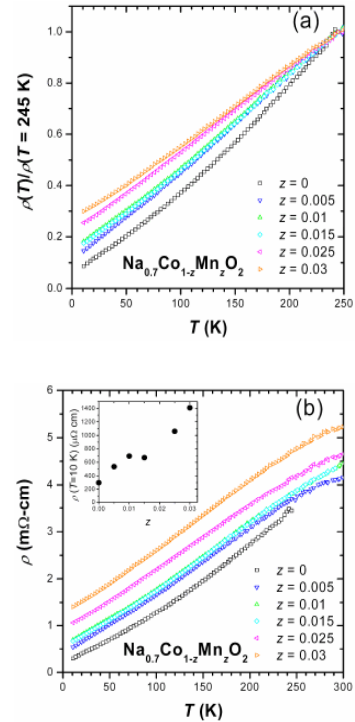

Fig. 4 Both (a) normalized $\rho(T)/\rho(T=245 K)$ and (b) $\rho(T)$ for $Na_xCo_{1-z}Mn_zO_2$ indicate an increase in $\tau_{imp}^{-1}$ due to Mn doping. In (b) data of $z$ = 0.01, 0.015, 0.025 are adjusted within a factor of 25% to simulate the Matthiessen's rule. Inset shows an increase in $\rho(T = 10$ K) with $z$.

Finally, $C(T)$ experiments reveal persuasive evidence supporting the coexistence model. In Fig. 5(a), $C(T)/T$ of $Na_xCoO_2 \cdot yH_2O$ is similar to that of Ref. [11] and the anomaly manifests the superconducting phase transition around 4.5 K. In Fig. 5(b), $C(T)/T$ of the doped sample $Na_xCo_{0.995}Mn_{0.005}O_2 \cdot yH_2O$ has a well defined anomaly consistent with $T_c$ measured by the $M$ measurements. More strikingly, there occurs another anomaly at lower $T$ (indicated by the arrow) which can be attributed to the second transition due to the nodal order parameter. It is noted that the transition temperature of this anomaly below 2 K is in quantitative agreement with the potential pair-breaking model as denoted by the dashed line in Fig. 3.

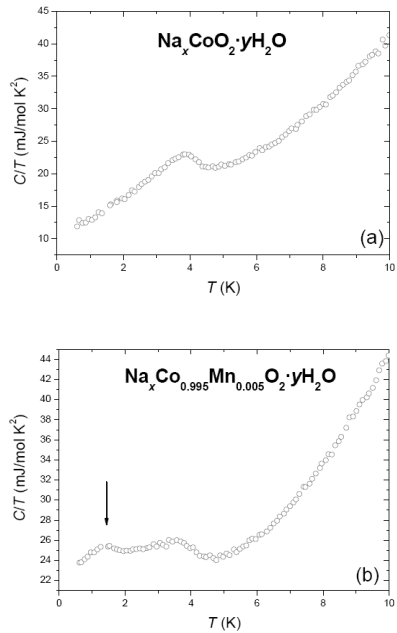

Fig. 5 $C(T)/T$ of (a) $Na_xCoO_2 \cdot yH_2O$ and (b) $Na_xCo_{0.995}Mn_{0.005}O_2 \cdot yH_2O$. The arrow indicates the lower temperature phase transition in addition to that consistent with $T_c$ measured by the $M$ measurements.

To summarize, Mn impurities have been successfully doped into $CoO_2$ planes in $Na_xCo_{1-z}Mn_zO_2 \cdot yH_2O$. Spectroscopy evidence indicates that the doped Mn ions are located on the Co sites. The absence of the strong impurity effects on $T_c$ is inconsistent with the simple picture of a sign-changing order parameter. The model of coexistence of $s$-wave and the unconventional pairings reconcile the observed $T_c$ suppression rate with other experiments. Two transitions attributed respectively to $s$-wave and the unconventional pairings have been directly observed by experiments.

We are grateful to H. Sakurai and M. Ogata for indispensable discussions. This work was support by the National Science Council of Taiwan, under grants: NSC 94-2112-M-009-006, 94-2112-M-018-001, and 94-2112-M-110-010.

*Corresponding author. E-mail address: ago@cc.nctu.edu.tw

[1] K. Takada, H. Sakurai, E. Takayama-Muromachi et al., Nature **422**, 53 (2003).
[2] M. L. Foo, Y. Wang, S. Watauchi et al., Phys. Rev. Lett. **92**, 247001 (2004).
[3] I. I. Mazin and M. D. Johannes, Nature Phys. **1**, 91 (2005).
[4] For example, H. D. Yang, J.-Y. Lin, H. H. Li et al., Phys. Rev. Lett. **87**, 167003 (2001).
[5] H. Ikeda, Y. Nisikawa, and K. Yamada, J. Phys. Soc. Jpn **73**, 17 (2004).
[6] M. D. Johannes, I. I. Mazin, D. J. Singh et al, Phys. Rev. Lett. **93**, 097005 (2004).
[7] M. Mochizuki, Y. Yanase, and M. Ogata, Phys. Rev. Lett. **94**, 147005 (2005).
[8] K. Kuroki, S. Onari, Y. Tanaka et al., cond-mat/0508482.
[9] K. Ishida, Y. Ihara, Y. Maeno et al., J. Phys. Soc. Jpn. **72**, 3041 (2003).
[10] A. Kanigel, A. Keren, L. Patlagan et al., Phys. Rev. Lett. **92**, 257007 (2004).
[11] H. D. Yang, J.-Y. Lin, C. P. Sun et al., Phys. Rev. B **71**, 020504(R) (2005).
[12] Y. Kobayashi, H. Watanabe, M. Yokoi et al., J. Phys. Soc. Jpn. **74**, 1800 (2005).
[13] Y. Ihara, K. Ishida, K. Yoshimura et al., cond-mat/0506751.
[14] P. Monthoux and D. Pines, Phys. Rev. B **49**, 4261 (1994).
[15] K. Mizuhashi, K. Takenaka, Y. Fukuzumi et al., Phys. Rev. B **52**, R3884 (1995).
[16] S. K. Tolpygo, J.-Y. Lin, M. Gurvitchet al., Phys. Rev. B **53**, 12454 (1996).
[17] S. K. Tolpygo, J.-Y. Lin, M. Gurvitch et al., Phys. Rev. B **53**, 12462 (1996).
[18] J.-Y. Lin, S. J. Chen, S. Y. Chen et al., Phys. Rev. B **59**, 6047 (1999).
[19] G. Harań and A. D. S. Nagi, Phys. Rev. B **63**, 012503 (2000).
[20] W. A. Atkinson, P. J. Hirshfeld, A. H. MacDonald et al., Phys. Rev. Lett. **85**, 3926 (2000).
[21] J.-Y. Lin and H. D. Yang, Phys. Rev. B **65**, 216502 (2002).
[22] A. P. MacKenzie, R. K. W. Haselwimmer, A. W. Tyler et al., Phys. Rev. Lett. **80**, 161 (1998).
[23] I. P. Hong, J.-Y. Lin, J. M. Chen et al., Europhys. Lett. **58**, 126 (2002); H. D. Yang, H. L. Liu, J.-Y. Lin et al., Phys. Rev. B **68**, 092505 (2003).
[24] D. G. Liu, J. F. Lee, and M. T. Tang, J. Mol. Catal. A **240**, 197 (2005); L. S. Hsu, Y.-K. Wang, Y.-L. Tai et al., Phys. Rev. B **72**, 115115 (2005).
[25] H. D. Yang, J.-Y. Lin, H. H. Li et al., Phys. Rev. Lett. **87**, 167003 (2001)
[26] M. Yokoi, H. Watanabe, Y. Mori et al., J. Phys. Soc. Jpn **73**, 1297 (2004).
[27] Q. Feng, H. Kanoh, Y. Miyai et al., Chem. Mater. **7**, 1226 (1995).
[28] R. Chen, P. Zavalij, and M. S. Whittingham, Chem. Mater. **8**, 1275 (1996).

[29] J. W. Lynn, Q. Huang, C. M. Brown et al, Phys. Rev. B **68**, 214516 (2003).
[30] J. D. Jorgensen, M. Avdeev, D. G. Hinks et al., Phys. Rev. B **68**, 214517 (2003).
[31] K. Takada, K. Fukuda, M. Osada et al., J. Mater. Chem. **14**, 1448 (2004).
[32] C. J. Milne, D. N. Argyriou, A. Chemseddine et al., Phys. Rev. Lett. **93**, 247007 (2004).
[33] C. P. Sun, J.-Y. Lin, H. D. Yang et al., unpublished.
[34] A. A. Abrikosov, *Fundamentals of the Theory of Metals* (North-Holland, Amsterdam, 1988).